\documentclass[twocolumn,showpacs,preprintnumbers,amsmath,amssymb]{revtex4}

\usepackage{graphicx}
\usepackage{dcolumn}
\usepackage{bm}

\begin{document}


\title{The Generation, Evolution and Decay of Pure Quantum Turbulence: A Full Biot-Savart Simulation}

\author{Shoji Fujiyama}
\author{Akira Mitani}
\author{Makoto Tsubota}
\affiliation{Department of Physics, Osaka City University, Sumiyoshi-ku, Osaka 558-8585, Japan}
\author{D.\,I.~Bradley}
\author{S.\,N.~Fisher}
\author{A.\,M.~Gu\'enault}
\author{R.\,P.~Haley}
\author{G.\,R.~Pickett}
\author{V.~Tsepelin}
\affiliation{Department of Physics, Lancaster University, Lancaster, LA1 4YB, UK. }

\date{\today}

\begin{abstract}

A zero temperature superfluid is arguably the simplest system in which to study complex fluid dynamics, such as
turbulence. We describe computer simulations of such turbulence and compare the results directly with recent
experiments in superfluid $^3$He-B. We are able to follow the entire process of the production, evolution,
and decay of quantum turbulence. We find striking agreement between simulation and experiment and gain new
insights into the mechanisms involved.
\end{abstract}


\maketitle The tangle of similar quantized vortices in a pure coherent condensate provides a very simple
model system for studying turbulence, which is more amenable to simulation and subsequent detailed analysis
than its classical counterpart.  Two recent developments inform the present work.  First, quantum turbulence
was recently observed in superfluid $^3$He. This medium turns out to be particularly useful as there are a
range of very sensitive well-developed techniques which can be easily adapted for turbulence studies.
Secondly, advances in computing power and techniques now allow us to make full Biot-Savart simulations of
turbulence over relatively large volumes allowing direct comparison with experiment.

We present a simulation of grid turbulence in superfluid $^3$He-B in the zero temperature limit, where the
normal fluid fraction is negligible, and compare with recent experiments. The vorticity is initially
generated as a gas of uniform vortex loops.  We are able to follow the ensuing collision and recombination of
these loops to form a turbulent tangle. After stopping the loop generation we can also follow the turbulence
as it freely decays. The simulation encompasses the full range of length-scale evolution from initial loops
to a developed tangle with large scale structure which decays back to small scales via a Richardson-like
cascade, thus providing an ideal scenario for understanding various aspects of pure quantum turbulence. The
simulation reproduces many of the features observed experimentally and provides further insights into the
processes involved.

In the low temperature limit, the few remaining thermal excitations in $^3$He-B comprise a highly ballistic
dilute gas of quasiparticles. The quantized circulation $\kappa=h/2m_3$ around vortex cores gives a very
large cross-section for Andreev reflection of quasiparticles, which allows experimental studies of vortices using
well-developed quasiparticle detection techniques which are only possible in $^3$He-B \cite{fisherwires,bradleywires}. Crucially, Andreev
reflection has no significant effect on vortex dynamics so the normal fluid can be disregarded, which greatly
simplifies the description and reduces computation times.

The simulation is directed towards recent experiments on a vibrating grid in $^3$He-B
\cite{gridrings,griddecay,gridresp,gridfluc,gridturb}. At low velocities the grid is found to emit a cloud of
independent vortex rings \cite{gridrings}. Computer simulations show that in a transverse oscillating flow,
vortex lines with fixed ends can produce rings \cite{Hanninen04}. When the frequency matches the resonance
condition for exciting the fundamental Kelvin-wave mode, the line couples strongly to the flow and it
stretches and twists with increasing absorbed energy. Eventually it reconnects to produce a free ring with a
size comparable to the initial line length \cite{Hanninen04}. The ring propagates away and the remaining
pinned vortex  repeats the process. The ring production rate increases monotonically with the velocity
amplitude of the oscillating flow \cite{Hanninen04}. Therefore, a possible scenario is that many vortex lines
of various lengths are pinned to the grid mesh used in the experiments, and these will produce a cloud of
vortex rings. The subsequent evolution of the cloud of rings depends crucially on the ring density. At low
densities, the rings propagate away from the grid at their self-induced velocity. Above a critical density
the rings no longer remain independent but, in a cascade, collide, reconnect and form a vortex tangle which
disperses on much longer timescales.

To simulate the grid experiments, we follow the dynamics of vortex
rings injected into a simulation ``cell'' such that the left hand side of the cell represents the face of the
grid. The experimental grid has dimensions 5.1\,$\times$\,2.8\,mm and is made up of a mesh of 10\,$\mu$m
square wires separated by 50\,$\mu$m. The simulation cell is necessarily smaller owing to computing
restraints, and is a box of cross-section $A=200\,\mu $m$\times200\,\mu$m and length $L_0$=600\,$\mu$m, shown
in Fig.~1. In the transverse directions the cell has periodic boundary conditions, which effectively allows
for vorticity entering the simulation volume from other parts of the grid. The box length is comparable to
the distance between grid and detector in the experiments. The experimental grid is likely to emit a range of
ring sizes, limited by the mesh size of 50\,$\mu$m and controlled by the Kelvin-wave resonance condition at
the grid oscillation frequency of $\sim$1300\,Hz, corresponding to ring diameters of $\sim$\,5\,$\mu$m, confirmed by
measurements of their decay at higher temperatures by mutual friction \cite{ringdecay}. In the
simulations, vortex rings of diameter $D$=20\,$\mu$m are injected at the left-hand side of the cell at a
regular time interval $\tau_i$ but at a random position and at a random angle within a $\theta
\sim$20$^\circ$ cone around the forward direction. The rings travel at a self-induced velocity of
$v=4.6$~mm/s \cite{Schwarz85}. We assume that vortex lines always reconnect on collision, as suggested by detailed
calculations of the intersection process \cite{Koplik93}.

\begin{figure} \includegraphics[width=0.75\linewidth]{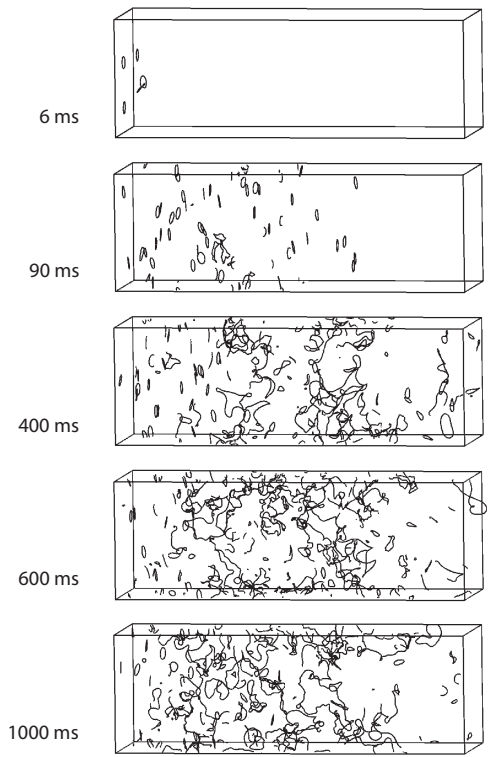} \caption{Simulation of quantum turbulence
formation. Each frame shows the vortex configuration at the labeled time. Rings injected from the left
quickly collide and recombine producing a vortex tangle which evolves on longer timescales.} \end{figure}

Our simulation methods \cite{Tsubota00} follow those of Schwarz \cite{Schwarz85}. A vortex line is
represented by a string of points along its core. The problem is greatly simplified at $T$=0 since, with no
normal fluid, each point moves with the local superfluid velocity which is calculated by full Biot-Savart
integration. (Although the order parameter of superfluid $^3$He is more complex, the equations which govern superfluid flow of $^3$He-B in low magnetic fields are identical to those of superfluid $^4$He, except that the vortices in $^3$He-B have larger, more complex, cores and the circulation quantum is a little smaller \cite{vollhardt}.) While there is no explicit decay process in the simulations, vorticity is lost in two ways.
First, vortices can escape through the open ends of the simulation cell, equivalent to the escape of
vorticity beyond the detector range in the experiment. Second, energy is lost when transferred to structures
smaller than the numerical space resolution set by the average point spacing on the filament (0.5\,$\mu$m in
our case). The justification of this cutoff procedure is discussed in detail in ref. \cite{Tsubota00}.

At a low injection rate ($\tau_i=5$\,ms), simulations confirm that the rings travel essentially
independently. Their relatively high speed results in a rapid loss of vorticity from the simulation cell when
the injection ceases. This corresponds to the rapid decay of the vorticity signal observed in the experiments
at low grid velocities \cite{gridrings}.

At higher ring injection rates, corresponding to higher grid velocities, we see very different behavior
as shown in Fig.~1 for $\tau_i = 1.5$\,ms. Here, the rings immediately start to collide and reconnect,
establishing a vortex tangle at the center of the box. This
corresponds to the behavior observed at high grid velocities in ref. \cite{gridrings}, namely quantum
turbulence which decays on much slower timescales. The simulations are consistent with a sharp
transition between the two regimes as observed experimentally.

We now turn to the transient behavior after switching on the ring generation. In Fig.~2 we plot the
build-up of the vortex line density (line length per unit volume) after commencing the ring injection. The upper plot shows line-densities
obtained in the simulations, averaged over the simulation volume. The highest curve corresponds to the
simulation of Fig.~1. At low ring injection rates, the rings propagate independently and fill the simulation
volume within the ring transit time $L_0/v \sim0.14$\,s. The steady state line density for independent vortex
rings is easily calculated. The injected vortex-ring flux along the forward direction is $(A\tau_i)^{-1}$.
Assuming that the rings travel independently and approximately in the forward direction, the
equilibrium line density for the rings is
\begin{equation}
L_{ring} \simeq\pi D/(A v \tau_i).
\end{equation}
This is shown by the horizontal lines in Fig.~2. At
higher injection rates the steady-state ring density occurs transiently, followed by a slower increase
as the tangle forms.

The lower part of Fig.~2 shows the experimental data for a range of grid velocities. At the lower velocities
it takes around a second for the full signal to develop,  the initial rise time being limited by the
mechanical response time of the grid resonator. At the higher velocities we see a further longer time
evolution coinciding with the onset of turbulence. Given the simple assumptions made, the correspondence
between simulation and experiment is striking. The line density increases as the tangle forms as this
sluggish object acts as an effective trap for incoming rings further increasing the line density. The balance
between the ring injection rate and the dissipation (see below) determines the equilibrium line density.

To estimate the timescale for developing the turbulent tangle, we note that the longest timescales are associated with the largest vortex structures which will have sizes approaching that of the
turbulent region $\mathcal{L}$. Assuming a Kolmogorov spectrum, the characteristic velocity on a length
scale $\mathcal{L}$ is $v(\mathcal{L}) \simeq C^{1/2} \epsilon^{1/3} (\mathcal{L}/2\pi)^{1/3}$ where $C\simeq
1.6$ is the Kolmogorov constant and the dissipation per unit mass can be written as $\epsilon = \zeta
\kappa^3 L^2$ \cite{Vinen}, where the dimensionless constant $\zeta \simeq 0.2$ is inferred experimentally
from the decay data \cite{griddecay}. We can then estimate the development timescale as
\begin{equation}
\tau \simeq \frac{\mathcal{L}}{v(\mathcal{L})} \simeq \frac{1}{\sqrt{C}} \left(\frac{2
\pi}{\zeta}\right)^{1/3}   \frac{1}{\kappa} \left(\frac{\mathcal{L}}{L}\right)^{2/3}.
\end{equation}
For the simulations, using $\mathcal{L} \simeq L_0=600$\,$\mu$m and $L \simeq 6 \times 10^8$\,m$^{-2}$ gives $\tau\sim
0.4$\,s,  and for the experiments with $\mathcal{L} \sim 1$\,mm and $L \simeq 1 \times 10^8$\,m$^{-2}$ gives $\tau\sim
2$\,s, in excellent agreement with the results in Fig.~2.

\begin{figure}
\includegraphics[width=\linewidth]{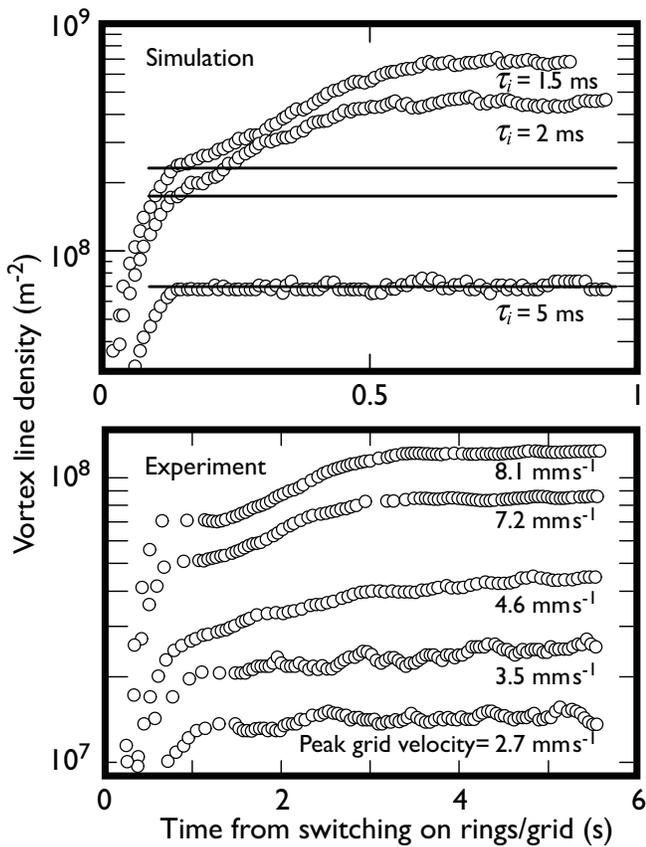} \caption{Upper figure: the mean vortex line density from the
simulations, for different ring injection-time intervals, steady-state line densities for free rings are
indicated by the horizontal lines. Lower figure: the vortex line density as measured for various grid velocities,
the tangle forms above $\sim 3.5$~mms$^{-1}$.}
\end{figure}

We may also estimate the critical vortex ring line density for the onset of turbulence. The
spread of ring speeds along the forward direction is $\sim v(1-\cos\theta) \sim v \theta^2/2$. To collide and
reconnect, two rings must approach within a ring diameter, thus the probability of any given ring
reconnecting per unit time is $\sim \pi D^2 \theta^2/(2A\tau_i)$. Multiplying by the cell transit time $\sim
L_0/v$ and the number of rings $\sim L_0/(v \tau_i)$ gives the total probability of a reconnection event
somewhere in the box, $p \sim ( \pi/2A)(D \theta L_0/\tau_i v)^2$. Once two rings have reconnected, the
larger fragment has a much lower velocity and is thus far more likely to collide with other rings, rapidly
leading to a vortex tangle. We may therefore estimate the condition for the onset of turbulence as $p \sim
1$, which gives us a critical ring injection time and a corresponding critical line density
\begin{equation}
L_c\sim \frac{1} {L_0 \theta} \sqrt{ \frac{2\pi}{A} } .
\label{Lc}
\end{equation}
This is a lower bound since it assumes that turbulence sets in as soon as any two rings collide, while in
fact more collisions may be needed. Applied to the simulation, equation  \ref{Lc} predicts $L_c \sim 6 \times
10^7$~m$^{-2}$ whereas the simulations in Fig.~2 show turbulence starting between $L=7 \times 10^7$~m$^{-2}$
and $L=2 \times 10^8$~m$^{-2}$. Comparison with experiment is harder since the critical density will be
affected by several other factors such as the complex three dimensional geometry and the distribution of ring
sizes and emission angles. However the simple expression gives roughly the correct order of magnitude for the
critical line density, with $\theta \sim 10 ^\circ$ and $L_0 \sim 1$~mm.

\begin{figure}
\includegraphics[width=0.75\linewidth]{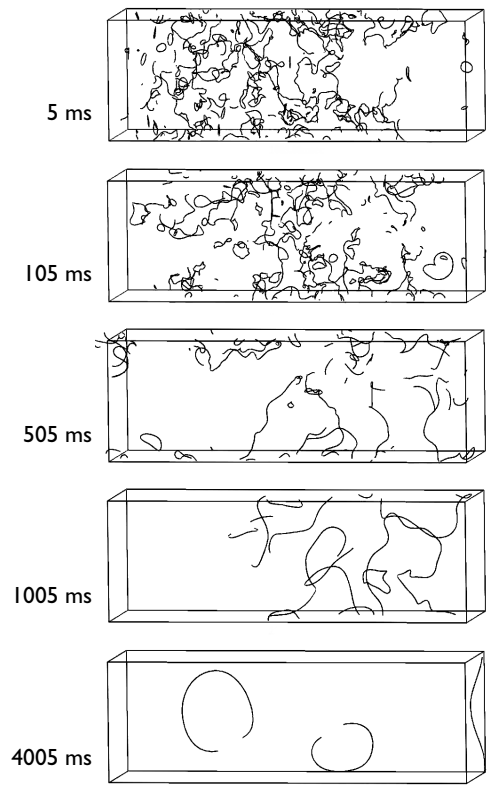} \caption{Simulation of the decay of vorticity.
The starting configuration is the last frame in Fig.~1. The frame labels indicate the time after cessation of ring
injection.}
\end{figure}

We now discuss the decay of the turbulence. Fig.~3 shows a continuation of the simulation of Fig.~1
after stopping the ring injection. By 0.1\,s the loss of the incoming rings leaves a clear
volume to the left with a slowly decaying tangle in the center. There is a small loss of vorticity through
the ends of the simulation cell but the majority of the loss is into features smaller than the computational
space resolution, which effectively gives a length scale below which vorticity is dissipated
as assumed in the Richardson cascade of classical turbulence.

\begin{figure} \includegraphics[width=\linewidth]{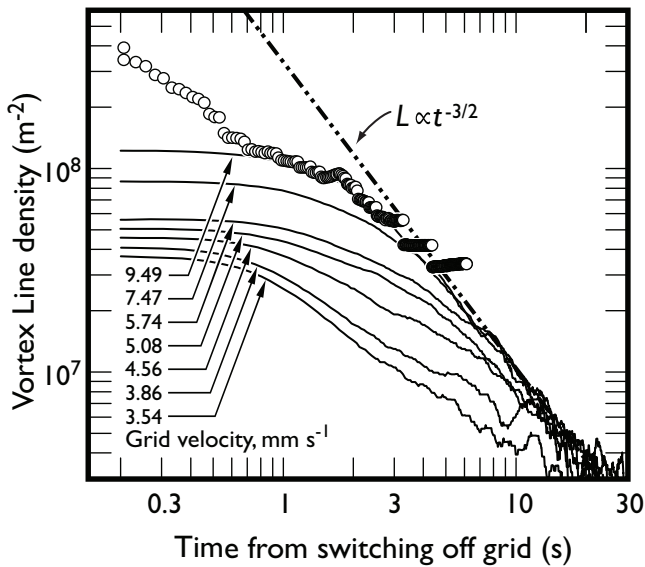} \caption{Decay of the vortex line density.
Circles show results from the simulation in Fig.~3. Lines show experimental data from ref. \cite{griddecay}.
Dashed line shows the expected late-time limiting behavior for a classical-like Richardson cascade.}
\end{figure}

In Fig.~4, we compare directly the decay of the line density from the simulation of figure 3 with the
experimental measurements at high grid velocities \cite{griddecay}. The late-time limiting behavior, $L\sim
t^{-3/2}$, shown by the dashed line in the figure, provides indirect evidence of a Richardson cascade
\cite{griddecay}, transferring energy from large to small length scales where dissipation ultimately takes
place. Similar behavior has been observed in superfluid $^4$He experiments \cite{Golov,Stalp} and in earlier
simulations \cite{TsubotaKolmogorov}. While in classical turbulence viscosity provides the decay mechanism,
turbulent decay in a superfluid at $T$=0 appears to be governed by the circulation quantum
\cite{griddecay,Golov,Stalp}, possibly from radiation by high-frequency Kelvin waves emitted by kinks left by
vortex reconnections \cite{Svistunov,Kelvinwaves,Vinen}. At late times the simulated line density runs into
problems of the limited simulation volume, however there is a remarkably close similarity to the experimental
data.  Given that the simulations simply remove energy from small length scales with no overt physical
mechanism, the result strongly suggests that, provided dissipation occurs only at small length scales, the
late time decay is insensitive to the precise dissipation mechanism.

In conclusion we have demonstrated that for a pure quantum fluid carrying turbulence we now have the
computational tools to undertake a full Biot-Savart simulation of the evolution of the vorticity over
significant volumes and time scales to make meaningful comparisons with experiments.  The agreement with the
observed behavior in the present case is quite startling.  The particular experiment chosen for this
comparison, grid turbulence, exhibits not only a Richardson-type cascade from larger to smaller scale
structures but also a reverse cascade from smaller to larger length scales as the initial gas of small
precursor vortex loops recombine to form larger and larger structures. The
simulations allow a detailed investigation of the various processes involved. Furthermore, the experimental
observation of a vortex-ring gas as a precursor to turbulence was completely unexpected but is easily
understood in the present simulations. In this paper we are limited to static figures. Many more details,
such as the influence of Kelvin waves produced by recombination events, are revealed by studying the full
time evolution (the full simulations can be seen as videos at
http://matter.sci.osaka-cu.ac.jp/~bsr/tsubotag). We expect that many further interesting results will arise
from more detailed analysis. For instance, we have recently been studying fluctuations in the vortex line
density which show a Kolmogorov-like, $-5/3$ power law, frequency spectrum as found experimentally
\cite{gridfluc,he4fluc}. Details will be reported elsewhere.

The ability to apply realistic simulations to experiments represents a significant milestone in the study of
quantum turbulence which will greatly enhance our interpretation of experiments and our detailed
understanding. We are also learning that quantum turbulence carries a strong resemblance to classical
turbulence and may therefore provide new insights into turbulence in general.

We acknowledge technical support from M.G.Ward and A.Stokes, and funding from: the Japanese JSPS; the Japanese MEXT; the UK EPSRC; the FP7 European MicroKelvin network; and the Royal Society.

\end{document}